\documentclass[twocolumn,amsmath,amssymb,floatfix,prl,superscriptaddress,showpacs]{revtex4}
\usepackage{amsmath,amssymb,natbib,bm,graphicx,url,epsfig,psfrag}
\usepackage[ansinew]{inputenc}

\newcommand{\be}{\begin{equation}}
\newcommand{\ee}{\end{equation}}
\newcommand{\bes}{\begin{equation}\begin{split}}
\newcommand{\ees}{\end{split}\end{equation}}

\begin{document}
\title{Electron Pair Resonance in the Coulomb Blockade}
\author{Eran Sela}
\affiliation{Department of Condensed Matter Physics, Weizmann
Institute of Science, Rehovot, 76100, Israel}
\author{H.-S. Sim}
\affiliation{Department of Physics, Korea Advanced Institute of
Science and Technology, Daejeon 305-701, Korea}
\author{Yuval Oreg}
\affiliation{Department of Condensed Matter Physics, Weizmann
Institute of Science, Rehovot, 76100, Israel}
\author{M.E.\ Raikh}
\affiliation{Department of Physics, University of Utah, Salt Lake
City, UT 84112, USA}
\author{Felix von Oppen}
\affiliation{Institut f\"ur Theoretische Physik, Freie Universit\"at
Berlin, Arnimallee 14, 14195 Berlin, Germany}
\date{\today}
\begin{abstract}
We study many-body corrections to the cotunneling current via a
localized state with energy $\epsilon_d$ at large bias voltages
$V$. We show that the transfer of {\em electron pairs}, enabled by
the Coulomb repulsion in the localized level, results in
ionization resonance peaks in the third derivative of the current
with respect to $V$, centered at $eV=\pm 2\epsilon_d/3$. Our
results predict the existence of previously unnoticed structure
within Coulomb-blockade diamonds.
\end{abstract}

\maketitle

\emph{Introduction.}---Current flow through a single localized
state (LS) coupled to metallic leads is a paradigm of quantum
transport through nanostructures, with applications to many
systems such as impurities embedded in tunnel barriers, quantum
dots, single-molecule junctions, or carbon nanotubes,  see,
\emph{e.g.}, Refs.\ \cite{CBreview,kouwenhoven98,
kouwenhoven01,molecular,bockrath}. Despite its simplicity, this
system exhibits a wide range of transport behaviors, including
resonant and sequential tunneling, cotunneling, and the Kondo
effect.

\begin{figure}[t]
\begin{center}
\includegraphics*[width=80mm]{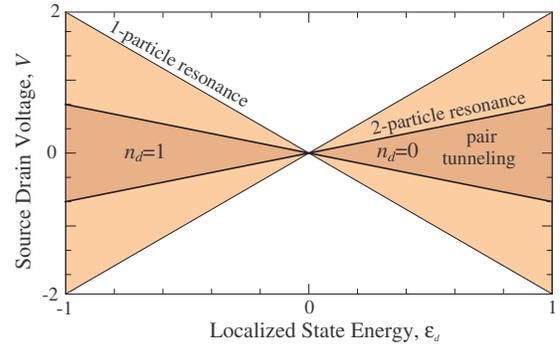}
\vspace {-.5cm}
 \caption{(Color online) Schematic stability diagram of
a single-level quantum dot. The thick lines within the shaded
Coulomb blockaded region are characterized by a resonance peak in
$d^3\mathcal{I}/dV^3$ due to opening of two-electron ionization. For
$eV$ below the threshold voltage $eV_c=2\epsilon_d/3$, tunneling of
electron pairs is a precursor effect to two-electron ionization.
\label{fig:1A}}
\end{center}
\vspace {-1.2cm}
\end{figure}

All of these regimes are captured remarkably well by a simple extension of
the Anderson impurity model
\begin{eqnarray}
\label{hamiltonian}
   && H= \sum_\sigma \epsilon_d d^\dagger_\sigma d_\sigma
   +U n_\uparrow n_\downarrow +\sum_{{\bf k}\sigma\alpha}
  \epsilon_{\bf k} c^\dagger_{{\bf k}\sigma\alpha}c_{{\bf k}\sigma\alpha}
  \nonumber\\
  && \,\,\,\,\,\,\,\,\,\,\,\,\,\,\,\,\,\,\,\,\,\,\,\,\,\,
  +\sum_{{\bf k}\sigma\alpha}
  \bigl[t_{\alpha}    d^\dagger_{\sigma}c_{{\bf
  k}\sigma\alpha}+t^*_{\alpha} c^\dagger_{{\bf k}\sigma\alpha}d_{\sigma}
  \bigr],
\end{eqnarray}
which describes tunneling of amplitude $t_\alpha$ between the
spin-degenerate LS of energy $\epsilon_d$ (with creation operator
$d^\dagger_{\sigma=\uparrow,\downarrow}$ and number operator
$n_\sigma=d^\dagger_\sigma d_\sigma$) and two leads $\alpha=L,R$
(with dispersion $\epsilon_{\bf k}$ and creation operator
$c^\dagger_{{\bf k}\sigma\alpha}$). For large on-site Coulomb
repulsion $U$, double occupation of the LS is suppressed and the
nature of transport depends on both $\epsilon_d$ (tunable by a
gate voltage $V_g$) and the bias voltage $V$. Within the shaded
areas of the stability diagram in Fig.\ \ref{fig:1A}, the average
occupation $n_d=n_\uparrow+n_\downarrow$ of the LS is close to
integer and current flow is suppressed by the Coulomb blockade. In
contrast, current can flow by sequential tunneling processes
outside the shaded areas, where the average occupation of the dot
is no longer integer. This picture of the Coulomb blockade has
been confirmed in numerous experiments performed on various
systems~\cite{CBreview,kouwenhoven98,kouwenhoven01,molecular,bockrath}.

It is the main point of this paper that even the minimal model of
\protect{Eq.~(\ref{hamiltonian})} predicts additional structure
{\em within} the Coulomb-blockaded region, emerging from {\em
two-electron} ionization of the LS at large biases. This
ionization process is an effect of many-body correlations, enabled
by the on-site Coulomb repulsion, which is much more robust than
the Kondo correlations emerging in the Kondo valley $n_d=1$ at low
temperatures and small voltages. Indeed, the fine structure due to
the two-electron ionization process exists in both Kondo and
non-Kondo valleys, as illustrated in \protect{Fig.~\ref{fig:1A}}.

The two-electron ionization requires biases beyond a threshold
voltage $V_c$, indicated by the thick black lines in
\protect{Fig.~\ref{fig:1A}}. Below the threshold voltage, correlated
{\em two-electron} transfers between the two leads constitute a
precursor effect to two-electron ionization. While the limit of the
Coulomb blockaded region is characterized by a resonance peak in the
differential conductance $d\mathcal{I}/dV$, we find that the onset
of two-electron ionization at $V_c$ is accompanied by a peak in
$d^3\mathcal{I}/dV^3$. Interestingly, the difference between both
resonance phenomena emerges solely from familiar Fermi liquid phase
space factors which appear in the two-electron ionization rate. One
important implication of this analogy is that the onset of
two-electron ionization is accompanied by anomalous temperature
sensitivity, even when $eV_c \gg T$, as is familiar for the boundary
of the Coulomb blockaded region.

\begin{figure}[t]
\begin{center}
\includegraphics*[width=85mm]{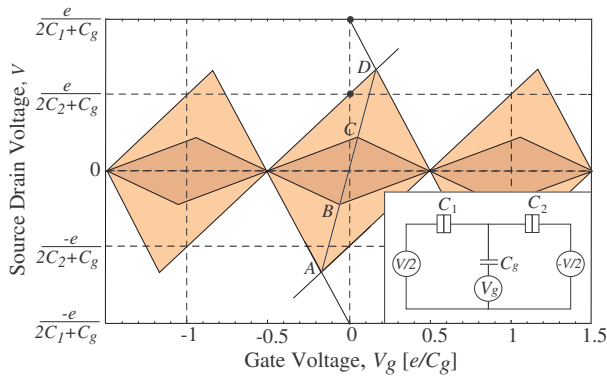}
\vspace{-.5cm}
 \caption{(Color online) Stability diagram of a
metallic quantum dot with vanishing level spacing. The threshold
lines for two-electron ionization can be determined by examining the
electrostatic energy of the circuit.  Graphically they are obtained
({\it e.g.}, for the central diamond) by rescaling the diagonal
$\overline{AD}$ by a factor $3$, so that
$\overline{AB}=\overline{BC}=\overline{CD}$. Inset: Equivalent
electric circuit for a metallic island. \label{fig:1D}}
\end{center}
\vspace{-.8cm}
\end{figure}

Most of our conclusions carry over to many-level quantum dots
(``metallic dots") where the stability diagram exhibits a sequence
of Coulomb diamonds, reflecting the step-wise population of the dot
with increasing gate voltage. This is depicted in
\protect{Fig.~\ref{fig:1D}} where we include the effects of
asymmetric capacitances between dot and electrodes.

In the remainder of the paper, we quantify the behavior of the
current near the two-particle threshold.

\emph{Two-electron ionization.}---Ionization by means of single
particle tunneling becomes energetically allowed when the source
chemical potential $eV/2$ is aligned with the LS, \emph{i.e.,} at
$eV=\pm 2\epsilon_d$. In contrast, the two-particle ionization
process, responsible for the predicted boundaries in the stability
diagram, is depicted in \protect{Fig.~\ref{fg:finalstate}(a)}. At
finite bias, an electron tunneling between the leads can suffer an
energy loss up to $eV$. Due to the on-site Coulomb repulsion, this
energy loss can be transferred to a second electron from the source
electrode, exciting it to energies up to $3eV/2$. Specifically, the
second electron can populate (and thus ionize) the LS once its
maximal energy exceeds $\epsilon_d$, \emph{i.e.}, for biases
exceeding the threshold voltage
\begin{equation}
   eV_c=2\epsilon_d/3.
   \label{eq:Vc}
\end{equation}
The predicted lines in the stability diagram originating from the
onset of two-electron ionization occur for $V=\pm V_c$. Thus, they
are located {\em within} the Coulomb blockaded region which
extends up to $eV=\pm 2\epsilon_d$.

Microscopically, the two-electron ionization process proceeds as
follows, cf.\ \protect{Fig.\ \ref{fg:finalstate}(a)}: (i) An
electron with energy $\epsilon_1$ from the source electrode (L)
enters the LS and (ii) tunnels into the state $E_1$ of the drain
(R). In the {\em same} process, (iii) a second electron with opposite spin
and energy $\epsilon_2$ tunnels from the source into the LS. The
amplitudes of the steps (i) and (iii) are proportional to $t_L$,
while the amplitude of step (ii) is proportional to $t_R^*$. Thus,
the resulting amplitude of two-electron ionization is given by
\begin{eqnarray}
\label{AMPLITUDE} A_{\epsilon_1\rightarrow
E_1}^{\epsilon_2\rightarrow \epsilon_d}=\frac{t_L^2t^{*}_R}
{(\epsilon_d-\epsilon_1)(E_1-\epsilon_1)}.
\end{eqnarray}
Following standard perturbation theory, the energy denominators are
given by the difference between the intermediate and initial
energies. In \protect{Eq.~(\ref{AMPLITUDE})}, we assumed a large
on-site Coulomb repulsion~$U$ so that there is no contribution from
virtual states with double occupation of the LS. If these states
were included, the corresponding terms would exactly cancel the
amplitude \protect{Eq.~(\ref{AMPLITUDE})} in the limit of vanishing
$U$. This makes it manifest that two-electron ionization is enabled
by the on-site Coulomb interaction.

Based on \protect{Eq.~(\ref{AMPLITUDE})\protect{ and energy
conservation, the two-electron ionization rate per spin, at $T=0$,
is
\begin{eqnarray}
\label{GAMMA}
\Gamma_{{\rm{ion}}}=\frac{\Gamma_L^2\Gamma_R}{(2\pi)^2}\int_{-\infty}^{eV/2}
d\epsilon_1\int_{-\infty}^{eV/2}d\epsilon_2\int_{-eV/2}^{\infty}dE_1\nonumber\\
\times
\frac{1}{(\epsilon_d-\epsilon_1)^2(\epsilon_2-\epsilon_d)^2}\,\delta(\epsilon_1+\epsilon_2-\epsilon_d-E_1),
\end{eqnarray}
where $\hbar=1$. Here $\Gamma_L=2 \pi |t_L|^2\nu$ and $\Gamma_R=2
\pi |t_R|^2\nu$ are the partial widths of the LS due to escape to
source and drain, respectively, and $\nu$ denotes the density of
states in the leads. Performing the integration over $\epsilon_2$,
we obtain
\begin{equation}
\label{GAMMA1}
\Gamma_{{\rm{ion}}}=\frac{\Gamma_L^2\Gamma_R}{(2\pi)^2}\int\limits_{-\infty}^{eV/2}
\!\!d\epsilon_1\int\limits_{-eV/2}^{\infty}\!\!\!dE_1
\frac{\theta\bigl(eV/2+\epsilon_1-E_1-\epsilon_d\bigr)}{(\epsilon_d-\epsilon_1)^2(E_1-\epsilon_1)^2},
\end{equation}
where $\theta(x)$ is the step function. Since $-eV/2 < E_1$ and
$\epsilon_1 < eV/2$, the argument of the $\theta(x)$ function is
negative for $\epsilon_d>3eV/2$, \emph{i.e.}, for $eV<eV_c$. In
contrast, for $0<V-V_c\ll V_c$, the integration regions for
$\epsilon_1$ and $E_1$ are restricted to $\epsilon_d-eV <
\epsilon_1 < eV/2$ and $-eV/2 < E_1 < eV - \epsilon_d$,
respectively. Since both regions are narrow, we find the threshold
behavior
\begin{equation}
\label{THRESHOLD} \Gamma_{{\rm{ion}}}=
\frac{9\Gamma_L^2\Gamma_R(V-V_c)^2}{32\pi^2 e^2 V_c^4}\theta(V-V_c)
\end{equation}
of the two-electron ionization rate $\Gamma_{{\rm{ion}}}$.

\begin{figure}[t]
\vspace{-.5cm}
\begin{center}
\includegraphics*[width=80mm]{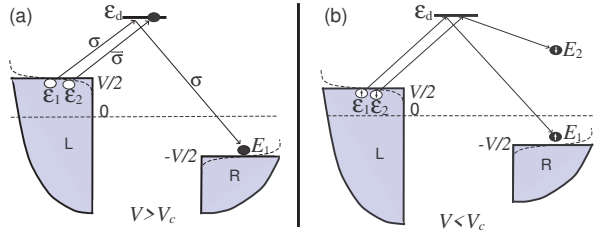}
\caption{(Color online) Schematic rendering of (a) two-electron
ionization of the LS, in which one electron tunnels from source to
drain while the other jumps into the LS, and (b) pair tunneling, in
which two electrons tunnel between source and drain. Both processes
are enabled by the on-site Coulomb repulsion suppressing double
occupation of the LS. } \label{fg:finalstate}
\end{center}
\vspace{-.8cm}
\end{figure}

It is crucial that energy exchange between electrons in the leads
does {\em not} require direct interaction between them. Instead,
this process is enabled by the finite Coulomb repulsion in the LS
alone. In this regard, the underlying physics of two-particle
ionization is similar to that of energy exchange between electrons
in a bulk metal, facilitated by a magnetic impurity
\cite{kaminski01}. Indeed, it is the non-zero on-site Coulomb
repulsion $U$ that ultimately generates the magnetic impurity \cite{Anderson}.
Curiously, similar many-body processes can also be enabled by the pairing interaction
in devices consisting of two Josephson junctions in series, where they lead to subgap
structure in the current \cite{Brink91}.

After entering the empty LS with rate $\Gamma_{{\rm{ion}}}$ by
two-electron ionization, the electron rapidly escapes into source
or drain electrode by single-electron tunneling. These
depopulation processes occur with rates $\Gamma_L$ and $\Gamma_R$,
respectively. Thus, the average occupation of the LS is governed
by the rate equation
\begin{equation}
 2\Gamma_{{\rm{ion}}}(1-n_{\uparrow})(1-n_{\downarrow})=(\Gamma_L+\Gamma_R)n_d.
\label{RATE}
\end{equation}
Here, the factor $2$ accounts for spin. \protect{Eq.~(\ref{RATE})}
yields $n_d \simeq
2\Gamma_{{\rm{ion}}}/\left(\Gamma_L+\Gamma_R\right)$. Since the
net charge transfer is $2e$ ($e$) when the electron tunnels out to
the drain (source) electrode, the ``two-electron ionization''
current $\mathcal{I}(V)$ between the leads becomes
\begin{eqnarray}
\label{CURRENT} \mathcal{I}(V)=e(2\Gamma_R+\Gamma_L)n_d\approx
2e\Gamma_{{\rm{ion}}}\frac{2\Gamma_R+\Gamma_L}{\Gamma_L+\Gamma_R}.
\end{eqnarray}
Due to $\Gamma_{{\rm{ion}}}$, the ionization current
$\mathcal{I}(V)$ also exhibits the threshold behavior
$\mathcal{I}(V)\propto (V-V_c)^2\theta(V-V_c)$.

Clearly, the ionization current, Eqs.~(\ref{CURRENT}) and
(\ref{THRESHOLD}), constitutes but a small fraction $\sim
\Gamma_L(V-V_c)^2/V_c^3$ of the cotunneling current $\sim
\frac{\Gamma_L \Gamma_R}{\epsilon_d^2} V$. Thus, it is an important
question how the threshold anomaly \protect{Eq.~(\ref{CURRENT})} can
be distinguished from the background cotunneling current.
\protect{Eq.~(\ref{CURRENT})} predicts that two-electron ionization
induces a jump in ${d^2\mathcal{I}}/{dV^2}$ located at $V=V_c$. We
now turn to a more careful analysis of this jump, focusing first on
the two-electron current below threshold, before deriving a general
interpolation formula.

{\em Two-electron current below threshold.}---For voltages below the
threshold, $V<V_c$, ionization of the LS is no longer possible by
two-electron processes. But two-electron processes can still excite
electrons in the leads to just below the energy of the LS. We will
now show that this constitutes a precursor effect to two-electron
ionization which contributes a logarithmically singular threshold
dependence to the differential conductance.

For large on-site Coulomb repulsion $U$, the two-electron process
below threshold proceeds microscopically as follows, cf.\ Fig.\
\ref{fg:finalstate}(b): ({i}) A spin-up electron from lead
$\alpha_1$ with energy $\epsilon_1$ enters the LS; ({ii}) the
electron tunnels out to state $E_1$ in lead $\alpha_1'$; ({iii}) a
spin-down electron from lead $\alpha_2$ with energy $\epsilon_2$
enters the LS and ({iv}) leaves into state $E_2$ in lead
$\alpha_2'$. The corresponding amplitude is
\begin{equation}
  A_{\epsilon_2 \rightarrow E_2}^{\epsilon_1 \rightarrow E_1}=\frac{t_{\alpha_1}t_{\alpha_2}t^*_{\alpha_1^\prime}t^*_{\alpha_2^\prime}
  }{(\epsilon_d-\epsilon_1)(E_1-\epsilon_1)(\epsilon_d-E_2)}+ (1 \leftrightarrow 2),
\label{amplitude}
\end{equation}
where the second term accounts for the four-step process described
above with the interchanged order (iii) $\mapsto$ (iv)$\mapsto$
(i) $\mapsto$ (ii). This results in a scattering rate
\begin{eqnarray}
\Gamma_{\alpha_{2}  \rightarrow \alpha'_{2} }^{\alpha_{1}
\rightarrow \alpha'_{1} } \!\!\!& = &\!\!\! 2 \pi \nu^4 \!\!\!
\int \!\! \prod_{i=1}^2\! \bigl[ d \epsilon_i d E_i
f(\epsilon_i-\mu_{\alpha_i}) \{ 1-f(E_i-\mu_{\alpha_i'}) \}\bigr] \nonumber \\
& \times & |A_{\epsilon_2 \rightarrow E_2}^{\epsilon_1 \rightarrow
E_1}|^2 \delta(\epsilon_1+\epsilon_2-E_1-E_2). \label{GOLDEN}
\end{eqnarray}
Here $f(\epsilon)=[e^{\epsilon/T}+1]^{-1}$ and $\mu_{L/R}=\pm eV/2$.
The resulting two-electron tunneling current contains two
contributions, $\mathcal{I } = I^{(1e)} + I^{(2e)}$, where $I^{(2e)}
= 2e \Gamma^{L \rightarrow R}_{L \rightarrow R}$ corresponds to
two-electron transfer between the leads, while $I^{(1e)}=e
\sum_{\alpha} (\Gamma^{L \rightarrow R}_{\alpha \rightarrow \alpha}
+ \Gamma^{\alpha\rightarrow \alpha}_{L \rightarrow R}) $ accounts
for one-particle transfer between the leads, accompanied by the
creation of a particle-hole excitation in one lead.

The crucial observation is that $\mathcal{I }$ is {\em singular} as
$V$ approaches $V_c$ from below. The singularity arises from the
domain $E_1\simeq -{eV}/{2}$, $E_2\simeq {3eV}/{2}$,
$\epsilon_1\simeq \epsilon_2 \simeq {eV}/{2}$. To see this, we first
note that in this domain, the amplitude $A_{\epsilon_2 \rightarrow
E_2}^{\epsilon_1 \rightarrow E_1}$ simplifies, $A_{\epsilon_2
\rightarrow E_2}^{\epsilon_1 \rightarrow E_1}\simeq
\frac{-t_{\alpha_1}t_{\alpha_2}t^*_{\alpha_1^\prime}t^*_{\alpha_2^\prime}}{(e
V_c)^2(\epsilon_d-E_2)}$. Using the Golden Rule
\protect{Eq.~(\ref{GOLDEN})}, and performing the integrals over
$\epsilon_1$, $\epsilon_2$, and $E_1$, we obtain for $e V_c \gg T$
\begin{eqnarray}
\label{SINGULAR} \mathcal{I}=\frac{2e}{h} \frac{\Gamma_L^2
\Gamma_R (\Gamma_R+\frac{1}{2}\Gamma_L)}{(2 \pi)^2 (e V_c)^4} \int
d E_2
\frac{1-f(E_2+e V/2)}{(\epsilon_d-E_2)^2} \nonumber \\
 \times f\Bigl(E_2-3eV/2\Bigr)
\Bigl[\left(\pi T\right)^2+\left(E_2-3eV/2\right)^2\Bigr].
\end{eqnarray}
Since  both $E_2$ and $\epsilon_d$ in the denominator of
\protect{Eq.~(\ref{SINGULAR})} are close to $3eV/2$, the remaining
integration yields the singular contribution
\begin{eqnarray}
\label{LOGARITHMIC} \frac{d\mathcal{I}}{dV}=\frac{2e^2}{h} \frac{3
\Gamma_L^2 \Gamma_R(\Gamma_R+\frac{1}{2} \Gamma_L)}{(2 \pi)^2 (e
V_c)^4} \ln \frac{eV_c}{\max \{ eV_c-eV, T \}}
\end{eqnarray}
to the differential conductance. The logarithmic singularity in the
two-electron tunneling current at $V_c$ signals the opening of the
two-particle ionization channel in Eq.~(\ref{CURRENT}) which is
lower order in the tunneling amplitudes and involves {\em real}
occupation of the LS.

The  appearance of $T$-dependence in Eq.~(\ref{LOGARITHMIC}) at
$eV_c\gg T$ resembles the behavior of the conductance near the onset
of sequential tunneling at $eV=\pm 2\epsilon_d$, cf.\ Fig.\ 1. In
fact, we find that the analogy between the onset of sequential
tunneling at $eV=\pm 2\epsilon_d$ and the onset of two-electron
ionization at $V=\pm V_c$ goes much further. The lines $eV=\pm
2\epsilon_d$ in the stability diagram separate transport regimes
with real occupation (sequential tunneling) and virtual occupation
(cotunneling) of the LS. Similarly, the lines $V=\pm V_c$ separate
regimes with real occupation (two-electron ionization) and virtual
occupation (pair-tunneling) of the LS. We now explore this analogy
on a quantitative level.

{\em Correspondence of one-electron and two-electron
ioni\-zation}.---We start by noting that Eqs.\ (\ref{CURRENT}) and
(\ref{LOGARITHMIC}) yield $d^2\mathcal{I}/dV^2 \propto
\theta(V-V_c)$ and $d^2\mathcal{I}/dV^2 \propto 1 / (V_c -V)$,
which are the familiar voltage dependencies of the
sequential-tunneling and cotunneling currents, respectively,
provided we make the replacement $eV_c \leftrightarrow
2\epsilon_d$. This suggests that the currents near the onsets of
sequential tunneling and two-electron ionization are related to
one another more generally by two voltage derivatives. To
establish this relation, although approximately, we incorporate
the lifetime broadening $\Gamma=\Gamma_L+\Gamma_R$ of the LS into
\protect{Eq.~(\ref{SINGULAR})}, and cast it into the form
\begin{eqnarray}
\label{APPROXIMATE} \mathcal{I}  \simeq\frac{2 e}{h}
\frac{\Gamma_L^2 \Gamma_R(\Gamma_R+\frac{1}{2}\Gamma_L)}{(2 \pi)^2
(e V_c)^4}\times \nonumber \\ \!\!\!\!\!\!\!\!\!\int
\frac{d\epsilon\left\{f(\epsilon-eV)-f(\epsilon+eV)\right\}}{\left[\epsilon-(\epsilon_d-eV/2)\right]^2+(\Gamma/2)^2}
\Bigl[(\pi T)^2+(\epsilon-eV)^2 \Bigr],
\end{eqnarray}
where $\epsilon \equiv E_2 - eV/2$. The finite lifetime provides a
physical cutoff of the singularity in Eq.\ (\ref{LOGARITHMIC}).
Most importantly, Eq.\ (\ref{APPROXIMATE}) captures processes
involving {\em both} virtual and real occupations of the LS,
\emph{i.e.}, it describes the \emph{two-electron resonance}.
Indeed, it can be easily verified that the above and
below--threshold limits, Eqs.\ (\ref{CURRENT}) and
(\ref{SINGULAR}), of the pair resonance are reproduced by Eq.\
(\ref{APPROXIMATE}). For $V \sim V_c$, Eq.\ (\ref{APPROXIMATE})
constitutes an approximate interpolation formula, due to the
attachment of an energy-independent width $\Gamma$ to the
two-particle resonance.

We compare Eq.\ (\ref{APPROXIMATE}) with a single-particle resonance
\begin{equation}
\label{COT} \mathcal{I}^{1PR}[V,\epsilon_d] = \frac{2 e}{h}
\Gamma_L \Gamma_R \int d \epsilon
\frac{f(\epsilon-\frac{eV}{2})-f(\epsilon+\frac{eV}{2})}{(\epsilon-\epsilon_d)^2+(\Gamma/2)^2}.
\end{equation}
The qualitative difference between the two expressions arises from
the appearance of the Fermi-liquid phase space factor $\bigl[(\pi
T)^2+ (\epsilon-eV)^2 \bigr]$ in the two-particle resonance
\protect{Eq.~(\ref{APPROXIMATE})}. This phase space factor can be
removed by taking two derivatives with respect to voltage of
\protect{Eq.~(\ref{COT})}. In this way, we find the relation
\begin{equation}
\label{RELATION} \frac{d^3 \mathcal{I}}{dV^3} \simeq \frac{\Gamma_L
(2\Gamma_R+\Gamma_L)}{(2 \pi)^2 e^2 V_c^4} \frac{d }{d V}
\mathcal{I}^{1PR}[2V , \epsilon_d-eV/2],
\end{equation}
with the explicit replacements $V \rightarrow 2V$ and $\epsilon_d
\rightarrow \epsilon_d-eV/2$. In view of the known properties of
the single-particle resonance, this result constitutes our
principal prediction. For $T \ll \Gamma$, Eq.\ (\ref{RELATION})
predicts a Lorenzian peak in $d^3 \mathcal{I}/dV^3$ inside the
Coulomb blockade diamond. Importantly, at $V =V_c$ both $d^3
\mathcal{I}/dV^3$ and $d^3 \mathcal{I}^{1PR}/dV^3$ have the same
order of magnitude $\sim {\Gamma^2}/{(h V_c^4)} $. These results
are illustrated in Fig.\ \ref{fg:Coulomb}. Note that, for $T \gg
\Gamma$, \protect{Eq.~({\ref{RELATION})} also predicts temperature
broadening of the peak in $d^3 \mathcal{I}/dV^3$.
\begin{figure}[t]
\begin{center}
\includegraphics*[width=80mm]{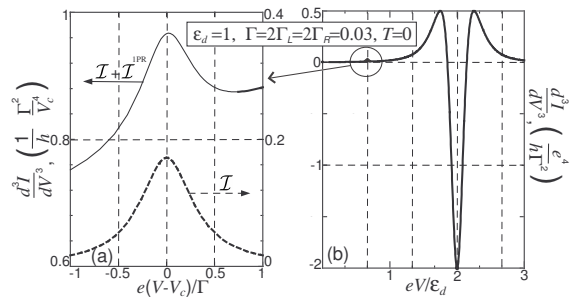}
\vspace{-.5cm} \caption{(a) The two-particle resonance induces a
peak (dashed line) in $d^3 \mathcal{I }/dV^3$ of width ${\rm max}
(T,\Gamma)$, centered at $V=V_c$. The full-line curve shows that the
two-particle resonance can be observed on top of the smoothly
varying single-particle background. (b) The single-particle
contribution to $d^3 \mathcal{I }^{1PR}/dV^3$ becomes singular at
$V=2 \epsilon_d$. \label{fg:Coulomb}}
\end{center}
\vspace{-1cm}
\end{figure}

{\em Metallic dots.---}Our predictions for transport via a single LS
also extend to metallic islands with essentially zero level spacing.
Transport through these islands can be modeled by the electric
circuit shown in the inset of \protect{Fig.~\ref{fig:1D}}. The
corresponding stability diagram includes a sequence of Coulomb
diamonds \cite{Devoret92}, cf.\ \protect{Fig.~\ref{fig:1D}}. It is
straightforward to see that for metallic dots, the boundaries of
two-electron ionization translate into a sequence of {\em inner
diamonds}, as shown in \protect{Fig.~\ref{fig:1D}}.

{\em Discussion and conclusion.---}Previously it was believed that
in the course of cotunneling through a LS, electrons from the source
arrive at the drain {\em one by one}. Here we demonstrated that
there exists a well-pronounced, although more delicate, transport
regime where two-electron processes contribute to the current. We
emphasize that this regime is captured by the standard Anderson
Hamiltonian \protect{Eq.~(\ref{hamiltonian})}.

Intriguingly, our reasoning is easily extended to regimes associated
with $N$-particle ionization of the LS ($N>2$). These induce
additional boundaries in the stability diagram Fig.\ \ref{fig:1A} at
even lower voltages $eV<eV_c^{(N)} = 2 \epsilon_d/(2N-1)$. A naive
estimate of the corresponding near-threshold behavior of the current
gives $\sim \theta(V-V_c^{(N)})[\Gamma^{2N-1}(eV-eV_c^{(N)})^{{2(
N-1)}}/(eV_c^{(N)})^{4(N-1)}]$. However, destructive interference
between different sequences of $N$-electron transitions might lead
to further reduction of the current.

Throughout this paper, we considered an empty LS at zero bias
(non-Kondo valley $\epsilon_d>0$). The analysis of the ionization
process of the occupied LS (Kondo valley $\epsilon_d<0$) is entirely
analogous, and differs only by the order of virtual transitions.

This work was supported in part by the DFG through Sfb 658 and Spp
1243 (FvO), DIP (FvO and YO), ISF and BSF (YO) as well as KOSEF,
KRF-2005-070-C00055, KRF-2006-331-C00118 (HSS). One of us (FvO)
gratefully acknowledges hospitality by the Weizmann Institute, made
possible by the EU - Transnational Access program
(RITA-CT-2003-506095).


\begin{thebibliography}{99}

\bibitem{CBreview} M.~A.~Kastner, Rev. Mod. Phys. \textbf{64}, 849 (1992).

\bibitem{kouwenhoven98} L.\ Kouwenhoven and C.\ Marcus, Physics World {\bf 11},
35 (1998).

\bibitem{kouwenhoven01} L.\ Kouwenhoven and L.I.\ Glazman, Physics World {\bf 14},
33 (2001).

\bibitem{molecular}J.\ Park {\em et al}., Nature (London) {\bf 417}, 722 (2002);
W.\ Liang {\em et al}., {\em ibid.} {\bf 417}, 725 (2002);
A.N.\ Pasupathy {\em et al}., Science {\bf 306}, 86 (2004).

\bibitem{bockrath} M.\ Bockrath {\em et al}., Science {\bf 275}, 1922 (1997).

\bibitem{kaminski01} A.\ Kaminski and L.I.\ Glazman, Phys.\ Rev.\ Lett.\ {\bf 86},
 2400 (2001).

\bibitem{Anderson} P.W.\ Anderson, Phys.\ Rev.\ {\bf 124}, 41 (1961).

\bibitem{Brink91} A.~Maassen~van~den~Brink, G.~Sch\"{o}n, and L.~J.~Geerlings, Phys.\ Rev.\ Lett.\ {\bf 67}, 3030 (1991); P.~Hadley {\em et al}., Phys.\ Rev.\ B {\bf 58}, 15317 (1998).

\bibitem{Devoret92} G.-L. Ingold and Yu. V. Nazarov, in \emph{Single Charge Tunneling}, edited
by H. Grabert and M. Devoret, NATO ASI, Ser.\ B, Vol.\ 294 (Plenum
Press, New York, 1992).


\end{thebibliography}
\end{document}